\documentclass[aps,prl,reprint,twocolumn,groupedaddress]{revtex4-1}
\usepackage{graphicx}

\begin{document}

%\title{Telecom-Heralded Single Photon Absorption by a Single Ion}
\title{Telecom-heralded single photon absorption by a single atom}

%\author{Andreas Lenhard}
%\author{Matthias Bock}
%\author{Stephan Kucera}
%\author{Jos\'e Brito}
%\author{Pascal Eich}
%\author{Philipp M\"uller}
%\author{Christoph Becher}
%\email[]{christoph.becher@physik.uni-saarland.de}
%\author{J\"urgen Eschner}
%\email[]{juergen.eschner@physik.uni-saarland.de}
%%\affiliation{Fachrichtung 7.2 (Experimentalphysik), Universit\"at des Saarlandes, Campus E2.6, 66123 Saarbr\"ucken, Germany}
%\affiliation{Experimentalphysik, Universit\"at des Saarlandes, Campus E2.6, 66123 Saarbr\"ucken, Germany}

% oder so:

\author{Andreas Lenhard}
\author{Matthias Bock}
\author{Christoph Becher}
\email[]{christoph.becher@physik.uni-saarland.de}
\affiliation{Quantenoptik, Fachrichtung 7.2, Universit\"at des Saarlandes, Campus E2.6, 66123 Saarbr\"ucken, Germany}
\author{Stephan Kucera}
\author{Jos\'e Brito}
\author{Pascal Eich}
\author{Philipp M\"uller}
\author{J\"urgen Eschner}
\email[]{juergen.eschner@physik.uni-saarland.de}
%\affiliation{Fachrichtung 7.2 (Experimentalphysik), Universit\"at des Saarlandes, Campus E2.6, 66123 Saarbr\"ucken, Germany}
\affiliation{Quantenphotonik, Fachrichtung 7.2, Universit\"at des Saarlandes, Campus E2.6, 66123 Saarbr\"ucken, Germany}

%\date{\today}

\begin{abstract}
We present, characterize, and apply a photonic quantum interface between the near infrared and telecom spectral regions. A singly resonant optical parametric oscillator (OPO) operated below threshold, in combination with external filters, generates high-rate ($>2.5\cdot10^6~{\rm s}^{-1}$) narrowband photon pairs ($\sim 7$~MHz bandwidth); the signal photons are tuned to resonance with an atomic transition in Ca$^+$, while the idler photons are at telecom wavelength. Quantum interface operation is demonstrated through high-rate absorption of single photons by a single trapped ion ($\sim 670~{\rm s}^{-1}$), heralded by coincident telecom photons.
\end{abstract}


\maketitle


\section{Introduction}

The vision of a quantum network integrates the concepts of classical communication with the potentialities of quantum physics, thereby opening up qualitatively new opportunities for data transmission and processing \cite{Kimble2008, Duan2010}. For long-range information broadcasting in quantum networks, the use of telecom fiber-compatible photons is highly desirable, due to the particular sensitivity of quantum information to loss. On the other hand, quantum information storage and processing makes use of optical transitions in atomic systems that often lie in the near-infrared (NIR) region. In order to interface atom-based stationary quantum bits with flying qubits realized by telecom photons, quantum frequency conversion techniques between the respective wavelengths may be used \cite{Zas12}. Another attractive approach, as presented in the following, is to interface the two spectral regions by (possibly entangled) photon pairs generated via spontaneous parametric down conversion (SPDC) \cite{Clausen2014, Saglamyurek2015, Fek13}.

Sources for photon pairs based on SPDC are available at telecom \cite{Fas04, Pomarico2012} and near infrared wavelengths \cite{Bao08, Haa09, Scholz2009, Wol11, Steinlechner2012}, and also based on integrated optics technology \cite{Fas04, Pom09, Kra13}. The SPDC process, however, is intrinsically too broadband ($\sim$100~GHz) for compatibility with atomic transitions ($\sim$10~MHz). Spectral shaping may either be achieved by external filtering \cite{Haa09, Piro2009}, or one may use an optical parametric oscillator (OPO) approach, by embedding the SPDC medium in an optical cavity. The latter was first demonstrated by Ou and coworkers \cite{Ou99} and also provides enhanced SPDC efficiency. OPO-based photon pair sources have been realized with bandwidths in the MHz range \cite{Scholz2007, Bao08, Scholz2009, Wol11, Monteiro2014}, and entangled photons compatible with atomic transitions in Cs \cite{Scholz2009} or Rb \cite{Bao08, Wol11} and at telecom wavelengths \cite{Pom09} have been generated. Most of these sources generate frequency degenerate photon pairs in doubly resonant cavities; spectral filtering of a non-degenerate OPO was introduced recently to generate photon pairs compatible with quantum memories and telecom networks \cite{Fek13}. 

On the atomic side, trapped Ca$^+$-ions provide a paradigmatic platform for implementing fundamental components of quantum information applications \cite{Schi13}. In the context of quantum networks, recent experiments showed the heralded absorption of single photons from a SPDC source by a single Ca$^+$ ion \cite{Pir11}, the manifestation of photonic entanglement in the absorption process \cite{Huwer2013}, bidirectional atom-photon state conversion \cite{Stute2012, Kurz2014, Casabone2015}, and the interconnection of ions via photonic channels \cite{Moehring2007, Schu13}. Here we report a photonic quantum interface based on an OPO source of non-degenerate photon pairs that bridges the NIR and telecom spectral regions and enables single-photon absorption in a Ca$^+$ ion heralded by a photon in the telecom range.

\section{Device Characterization}

The optical parametric oscillator (OPO) is based on a design originally developed for experiments in quantum frequency conversion at telecom wavelengths \cite{Zas11}. Its heart is a 30~mm long crystal of periodically poled, MgO-doped stoichiometric lithium tantalate, whose end facets are cut under $2^\circ$ angle and anti-reflection coated for pump, signal, and idler wavelengths. Its pump source is a frequency-doubled solid state laser system (532~nm, max. 10~W). Six equally spaced poling periods between $\Lambda_1=8.1$~$\mu$m and $\Lambda_6=8.6$~$\mu$m allow for OPO operation anywhere between 1202~nm and 1564~nm idler wavelength and 806--954~nm signal wavelength. The crystal is placed in a four-mirror bow-tie type ring cavity \cite{Zas10}. Three mirrors are coated for high reflectivity at 790--995~nm and high transmission at the pump and idler wavelengths; an output-coupling mirror with 97~\% reflectivity at the signal wavelength allows us to operate the OPO below threshold (2.8~W) as a SPDC-source of individual signal and idler photon pairs. For the experiments reported here we use 300~mW pump power, as a trade-off between high pair generation rate and low multi-pair probability. The OPO covers the telecom O-, E-, S- and C-band and simultaneously the NIR region where transitions of atoms and atom-like systems such as Ca$^+$ (854~nm), Cs (894~nm), InAs/GaAs quantum dots (900-940~nm) \cite{Mat12, Ulh12}, or solid state quantum memories (883~nm) \cite{Cla11} are found.

\begin{figure}[bth]
	\includegraphics[width=0.43\textwidth]{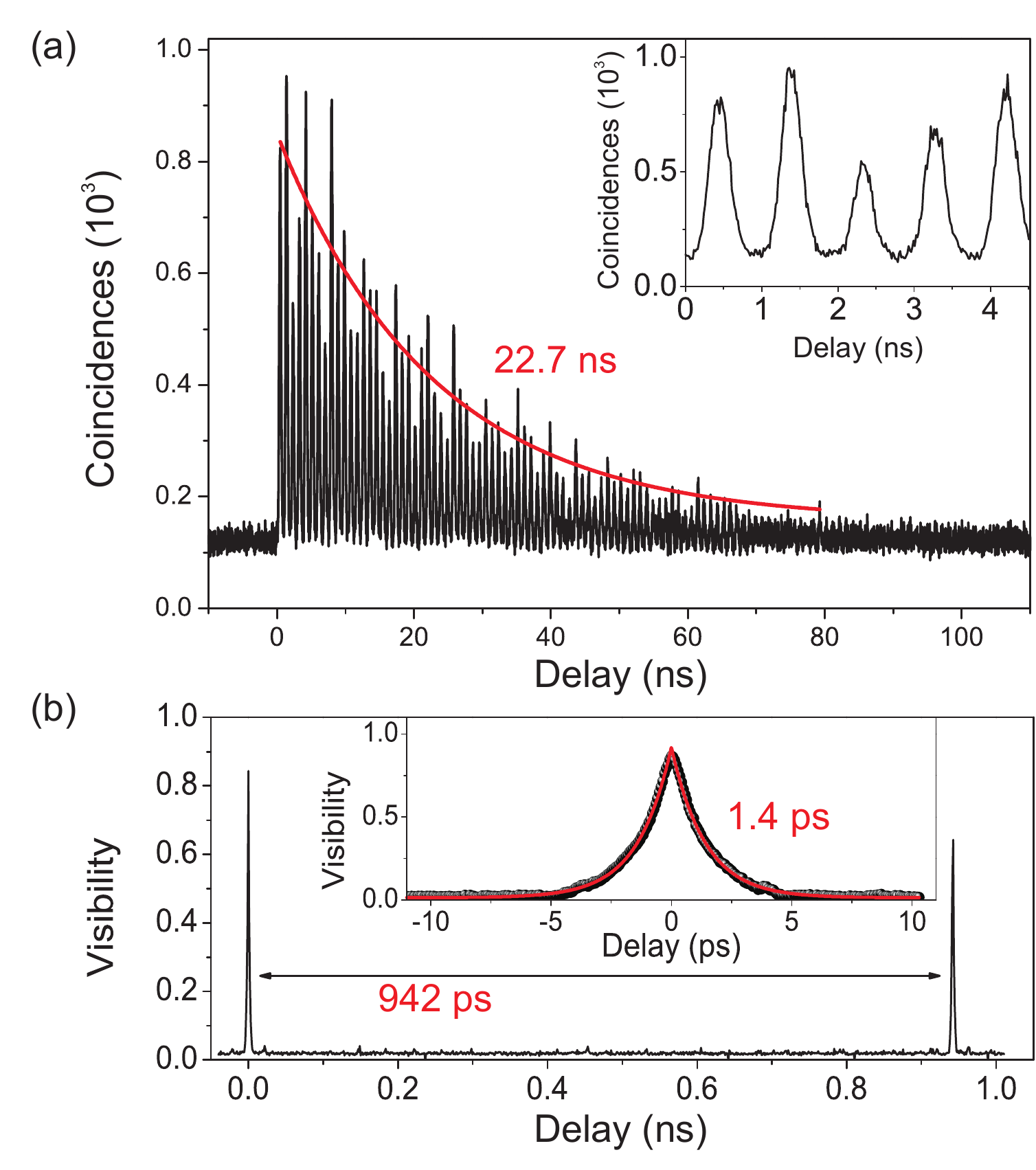}
	\caption{Temporal characteristics of the photon pairs: a) Time correlation between the heralding idler and the signal photons. The idler photon is detected by an InGaAs avalanche photo diode (APD) and used as a start trigger for time-tagged single photon counting of the signal photon which is detected by a Si-APD. The inset is a close-up of the data. b) First order coherence function of the signal photons. Inset: high resolution scan of the peak at zero delay, with an exponential fit function. No noise subtraction is applied. The maximum visibility of 0.9 is mainly limited by dark counts. The detection rate in both plots was $1.7\cdot10^4$~cts/s. \label{fig:TempCoherence}}
\end{figure}

We expect the spectrum of the signal photons to be a convolution of the Airy-function of the cavity with the phase matching function of the OPO crystal. While the free spectral range of the cavity, $\sim 1$~GHz, is not resolved with our spectrometers, it is reflected in the temporal shape of the photon wave packet: if a photon pair is generated in the nonlinear crystal, the idler photon will leave the cavity immediately while the signal photon is likely to be reflected at the output coupling mirror. As shown in Fig.~\ref{fig:TempCoherence}a, the temporal correlation between idler and signal photon detection reveals the cavity round-trip and the cavity ring-down. An exponential fit to the envelope yields a cavity decay time of $22.7\pm3.6$~ns, corresponding to a linewidth of $7.2\pm1.1$~MHz. The well-separated peaks are spaced by the cavity round-trip time of $939\pm4$~ps. The width of these lines is resolution-limited due to detection jitter in the APDs.

To gain more insight into the photons' spectral properties we measure the first-order coherence function of the signal field using a Michelson interferometer. Figure~\ref{fig:TempCoherence}b shows the interferogram (visibility as a function of delay) around zero delay. In the long-range scan (main graph) peaks appear with a spacing of $942\pm1$~ps, confirming the previous result for the round trip time of the OPO cavity. The inset shows a high-resolution scan; a two-sided exponential fit yields 1.4~ps coherence time, which corresponds well to the measured spectral width of 275~GHz (FWHM). This width is mainly determined by the phase matching conditions of the nonlinear process. The contrast and the comb structure visible in the temporal and coherence function are in good agreement with the theoretical prediction \cite{Her08}.

According to these results, the width of a single mode from the comb-like spectrum is 7.2~MHz, already compatible with atomic resonances. To reduce the background from other, non-resonant modes, the resonant single line is extracted by narrowband filtering of the coincident idler photon that serves as a herald for its atom-resonant partner \cite{Scholz2007, Haa09, Scholz2009, Wol11, Kra13, Fek13}. As we target a transition in Ca$^+$ at 854~nm, the corresponding idler wavelength is at 1411~nm. We use a narrowband tunable filter system (AOS GmbH) based on two cascaded fiber Bragg gratings (FBG). This device has a Lorentzian transmission window of 1.56~GHz (FWHM) and a rejection band exceeding by far the phase matching spectrum. The filter system has 20~\% maximum transmission, including coupling loss between the OPO and a single mode fiber. As the filter width exceeds the OPO free spectral range, neighbouring modes are also partially transmitted and will contribute a background of heralds that are not coincident with a resonant photon. From numerical calculations we expect a signal to background ratio around~1. 

\begin{figure}[tbh]
	\includegraphics[width=0.45\textwidth]{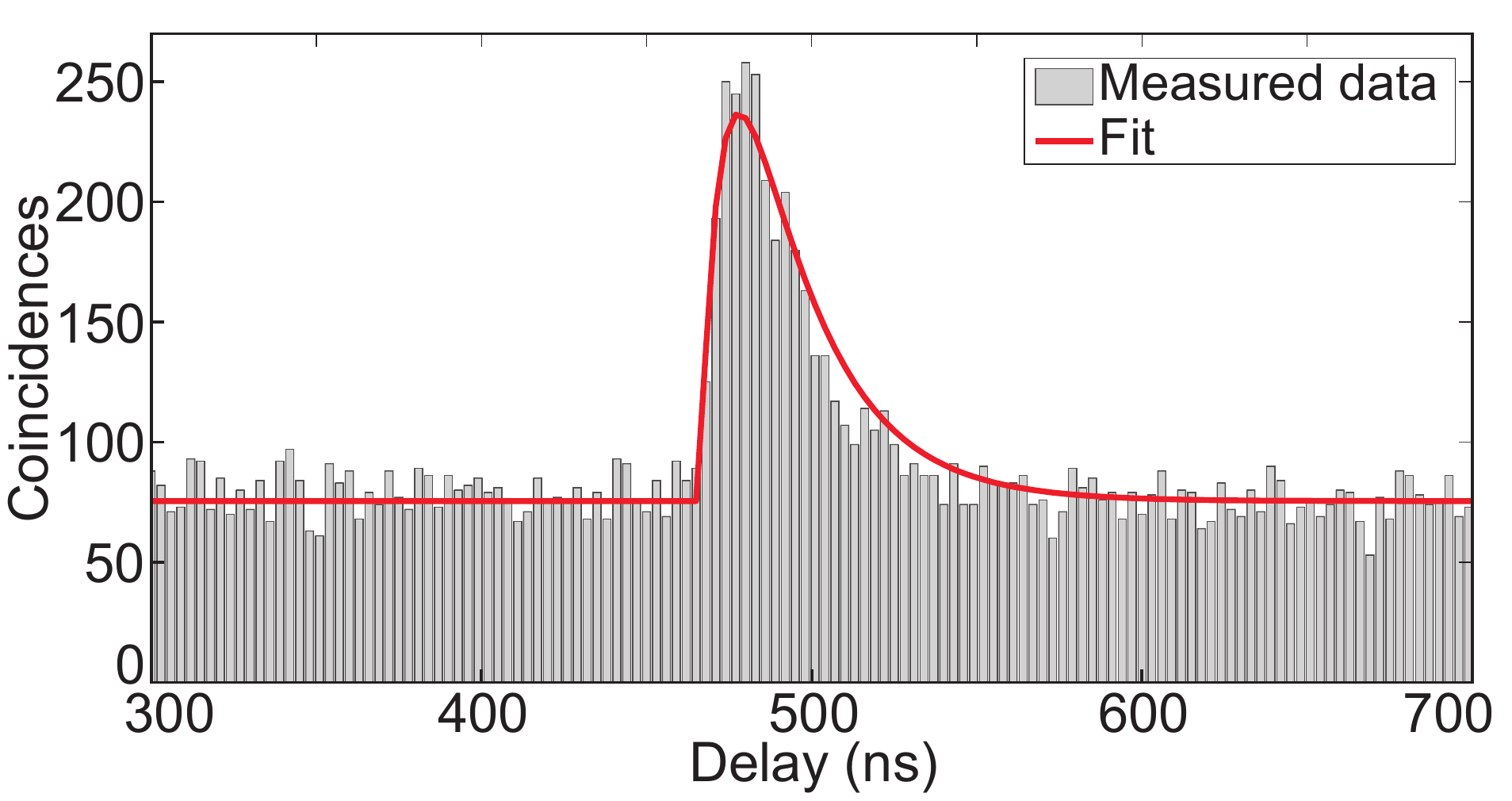}
	\caption{Correlation measurement of narrowband idler (filtered by FBG, start) and signal (cavity filter, stop) photons. Time bin size is 3~ns, total measurement time is 30~minutes. \label{fig:CorrelationFilter}}
\end{figure}

To characterize the narrowband photon pairs we also filter the signal photons down to a single mode. The filter system is described in \cite{Haa09, Piro2009}. It consists of two cascaded Fabry-P\'erot cavities with different free spectral ranges resulting in a transmission bandwidth of 22~MHz, tailored to match the linewidth of the D$_{5/2}$--P$_{3/2}$ transition of the $^{40}$Ca$^{+}$-ion at 854~nm. Both filter cavities are actively stabilized to a laser which is resonant with the atomic transition. Transmitted photons are detected by a Si-APD. Thus this filtering system serves to emulate the calcium transition of interest by detecting only photons from the resonant single mode of the OPO spectrum.
%{\bf The OPO is passively stable but we observed frequency fluctuations and drifts in the 100~MHz range. We continuously scan the OPO cavity length via the piezo mounted mirror and thus tune the photon frequency periodically over the filter window.} 
The result of the correlation measurement between filtered idler and signal photons is shown in Fig.~\ref{fig:CorrelationFilter}. The data is well described by a convolution of two exponential decays, as shown. The first time constant is the filter cavity decay time, 7.0~ns, while the second time constant corresponds to the width of the photon wave packet, $22.7$~ns, as determined above. Since the loss factors between the OPO and the two detectors were not fully characterized in this measurement, we defer the analysis for the generated pair rate until the next section. 

\section{Heralded Absorption}

\begin{figure}[bt]
	\includegraphics[width=0.48\textwidth]{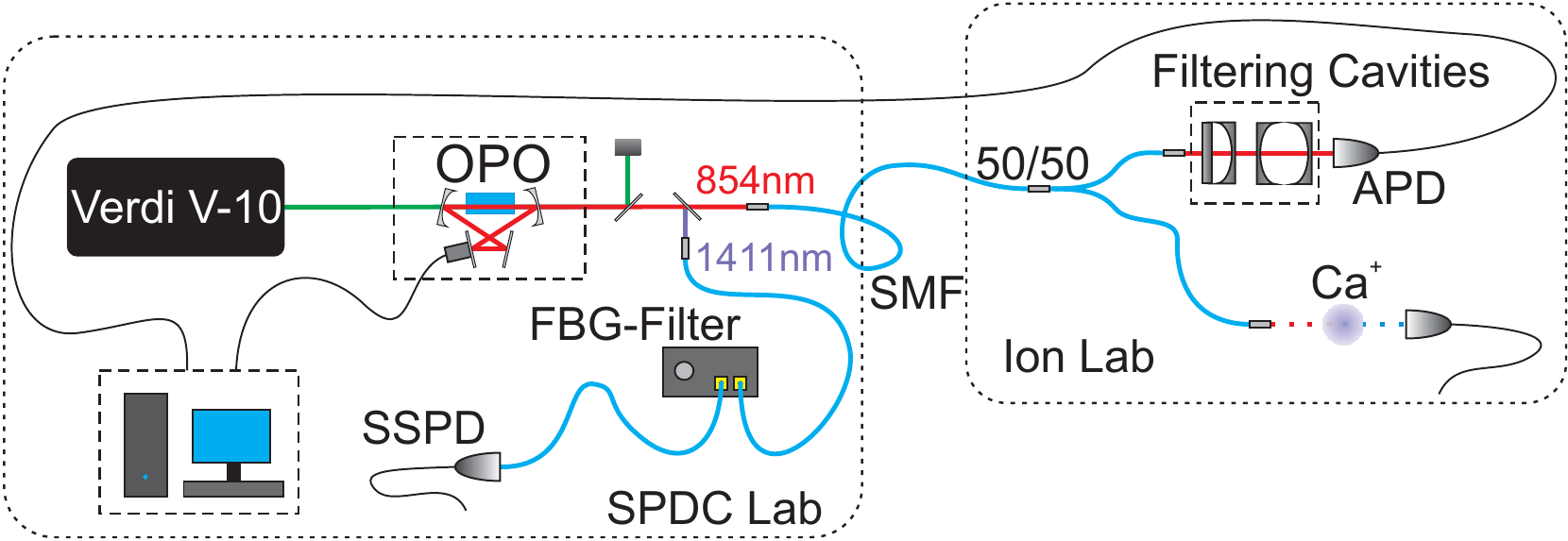}%
	\caption{Schematic drawing of the experimental setup. The two labs (dashed boxes) are separated by 90~m. For details see text. \label{fig:Setup}}
\end{figure}

A sketch of the full experimental setup is displayed in Fig.~\ref{fig:Setup}. The OPO and the ion trap are located in different labs connected via a 90~m single mode fiber (SMF) link. At its output, a fiber beam splitter guides half of the 854~nm idler photons to the ion and the other half to the cavity filtering system described above. The detection rate behind the filter is utilized in a feedback loop to stabilize the center frequency of the OPO against fluctuations and drifts due to environmental conditions. The photons sent to the ion do not pass any filter. 

\begin{figure}[htb]
	\centering
	{\includegraphics[width=0.45\linewidth]{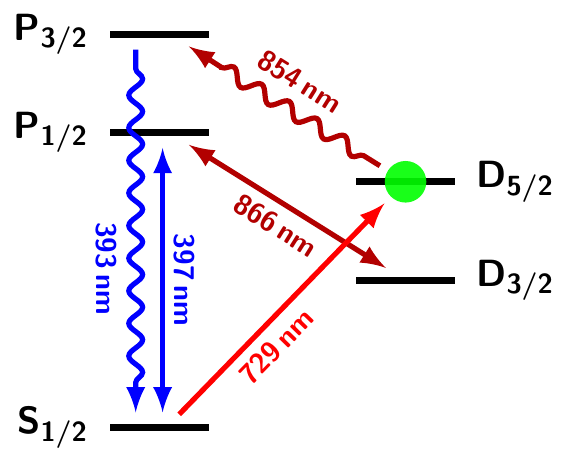}}
	\caption{Relevant energy levels and transition wavelengths of $^{40}$Ca$^+$. The green dot denotes the initial state for the single-photon interaction.}
	\label{fig:LevelScheme}
\end{figure}

The calcium ion is prepared for single-photon absorption in the meta-stable D$_{5/2}$ fine-structure manifold (lifetime 1.17~s), see Fig.~\ref{fig:LevelScheme}. Absorption of an 854~nm photon on the D$_{5/2}$ to P$_{3/2}$ transition, followed by spontaneous decay to S$_{1/2}$, induces a quantum jump of the ion signalled by the onset of atomic fluorescence (at 397~nm wavelength, see \cite{Pir11}). Single-photon absorption is detected with about 94\% efficiency, set by the branching fraction for decay of the P$_{3/2}$ level to S$_{1/2}$.  

In detail, the experimental sequence proceeds as follows: in the initial 60~$\mu$s, the ion is optically pumped (with 99.6~\% efficiency) into the $|{\textrm S}_{1/2}, m=-\smash{\frac{1}{2}}\rangle$ Zeeman sub-level of the ground state. Afterwards, a 2.2~$\mu$s long narrowband laser pulse at 729~nm transfers 99.2~\% of the population to the $|\text{D}_{5/2}, m=-\smash{\frac{5}{2}}\rangle$ state. Three additional pulses clean out any remaining S$_{1/2}$ population to other D$_{5/2}$ sub-levels; in total, preparation in D$_{5/2}$ happens in 11~$\mu$s with 99.99~\% probability. Then the cooling lasers at 397~nm and 866~nm, driving the transitions S$_{1/2}$--P$_{1/2}$ and D$_{3/2}$--P$_{1/2}$, respectively, are switched on, and the ion is exposed to the photons from the OPO for 7~ms. Ion-photon coupling is optimised by the use of a high-aperture laser objective (HALO). The fluorescence at 397~nm, whose onset serves as evidence for the absorption, is collected by two HALOs and detected by two photomultiplier tubes (PMT) with 1.5\% overall efficiency, resulting in 4.3~$\mu$s temporal resolution for the detection of a quantum jump. After the interaction time the sequence starts again. The probability for spontaneous decay from D$_{5/2}$ to S$_{1/2}$ during the interaction time is 0.6~\%. 

\begin{figure}[tb]
	\includegraphics[width=0.45\textwidth]{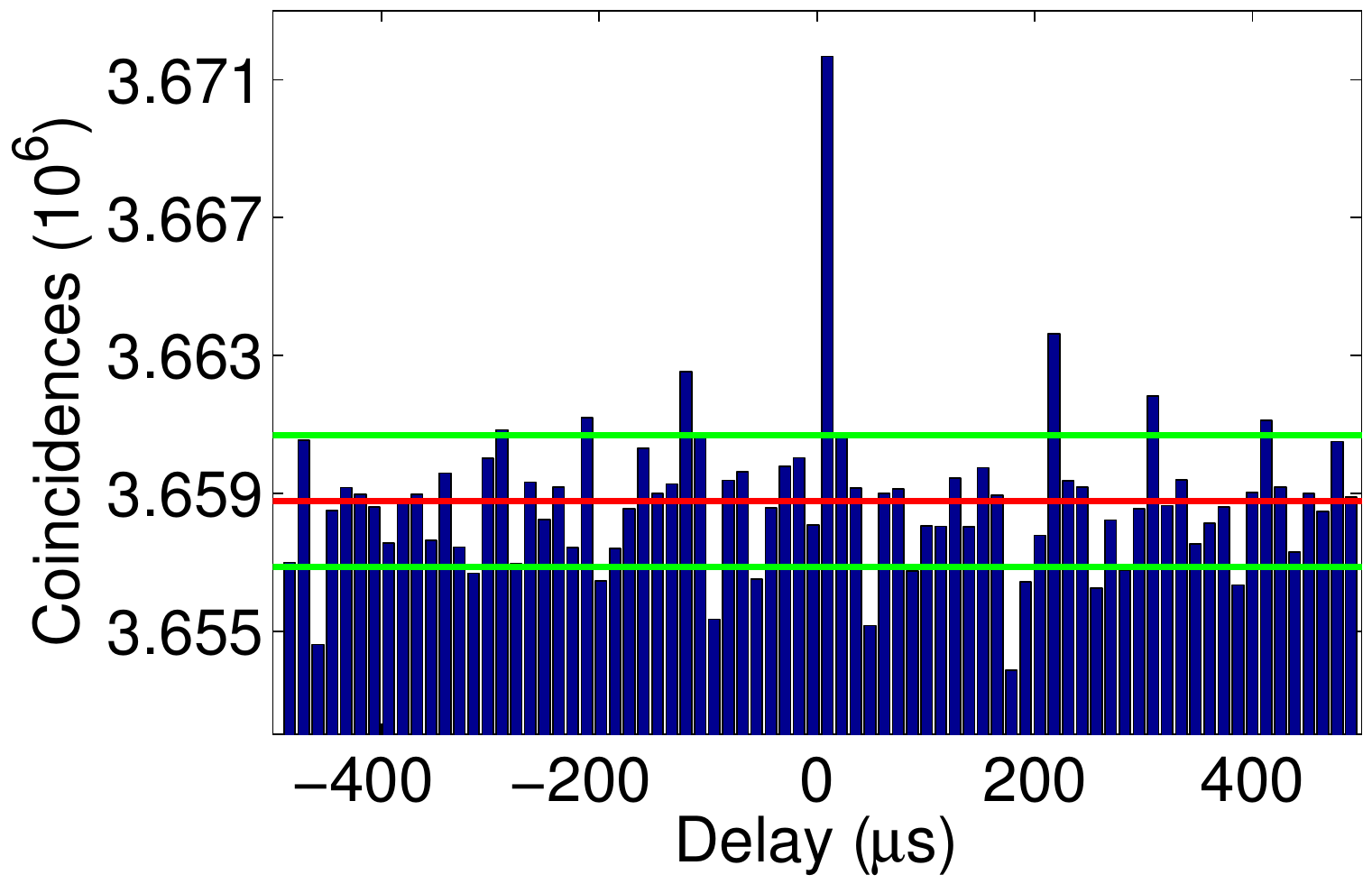}
	\caption{Coincidences of quantum jumps and filtered telecom photons. The red line indicates the background level, and the green lines the Poissonian noise interval of the background. Time bin size is 13~$\mu$s, total measurement time is 310 min. \label{fig:QuantumJumps}}
\end{figure}

In parallel the telecom idler photons transmitted by the narrowband FBG filter are detected by a superconducting single photon detector (SSPD).  
%(SSPD, Single Quantum Eos X10). 
Both the 397~nm fluorescence photons and the detected telecom heralds are recorded with time-tagged photon counting electronics. 
%(Picoquant Pico Harp 300). 
The quantum jumps are found in the post-processing of the data \cite{Piro2015} and then correlated with the detection times of the telecom photons. The result is shown in Fig.~\ref{fig:QuantumJumps}. For the coincidence peak around zero delay we find a signal-to-noise ratio of 6.7. The result clearly manifests the temporal correlation between the quantum jumps and the telecom photons, i.e., the heralded absorption of single OPO photons by the single ion. The background of uncorrelated events is mainly produced by photons with lost partners, including ion-resonant photons that are not absorbed. Dark counts play a negligible role. From the observed peak values of the absorption rate, 680~s$^{-1}$ (see supplement \cite{Sup}), and the detection rate on the Si-APD, 6,500~s$^{-1}$, and taking into account the known losses, we estimate the resonant pair generation rate of the OPO to be at least $2.5 \cdot 10^6$~s$^{-1}$. Divided by the OPO pump power this results in 8,400 pairs$(\rm s ~ mW)^{-1}$; this value exceeds the performance reported for other state-of-the-art OPO SPDC devices \cite{Bao08, Fek13, Scholz2009}. The absorption probability per resonant photon sent into the ion trap is $\sim 2\cdot 10^{-3}$. An analysis based on signal and idler detection rates confirms these numbers. Details are provided in the supplementary material \cite{Sup}.

\section{Single Photon Spectroscopy}

\begin{figure}[th]
	\includegraphics[width=0.4\textwidth]{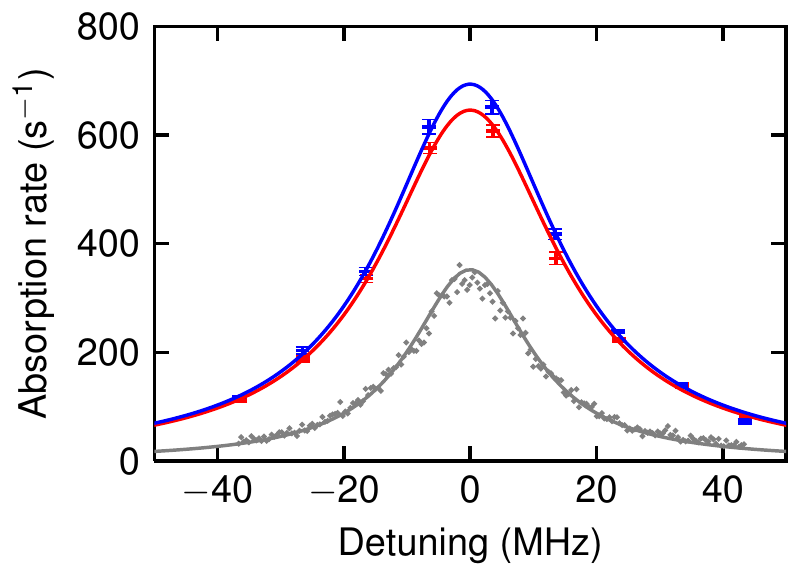}
	\caption{Spectroscopy of the atomic line with single photons from the OPO. Absorption rate as a function of OPO detuning. In order to compensate for frequency drifts of the OPO, individual data sets are recorded for detuning to higher (blue) and lower (red) frequencies; Lorentzian fits are shown. For comparison, a laser excitation spectrum is also shown (grey points and line, not to scale). \label{fig:spectrum}}
\end{figure}

To demonstrate quantum interface operation with our photon pair source we use the narrowband photons from the OPO for spectroscopy of the atomic transition. In particular, we study how the lifetime of the metastable D$_{5/2}$ level is shortened by the single-photon excitation, compared to its value $\tau_{\rm sp}=1.17$~s in the absence of resonant photons. Controlled tuning of the OPO is effected by varying the center frequency of the cavity filters to which the OPO is stabilized via a side-of-fringe lock. We measure the effective lifetime $\tau_{\rm eff}(\Delta)$ of the D$_{5/2}$ state vs.\ the detuning $\Delta$ as follows: the ion is prepared as described in the previous section, and the time between the preparation and the quantum jump (onset of fluorescence) is measured. The histogram of this time distribution is used for a Bayesian estimation of the decay probability, which is then fit by an exponential function yielding the effective lifetime, $\tau_{\rm eff}$. The absorption rate is derived as $R_{\rm abs} = (\tau_{\rm eff}^{-1} - \tau_{\rm sp}^{-1})/0.94$, which corrects for spontaneous decay from D$_{5/2}$ to S$_{1/2}$ and for the fraction of undetected absorptions (6\%) that lead to decay back to D$_{5/2}$. The result is shown in Fig.~\ref{fig:spectrum}. 

From the two Lorentzian fits to the data we find a mean peak absorption rate $R_{\rm abs}=670~{\rm s}^{-1}$, and a mean linewidth of $33.6\pm0.6$~MHz (FWHM). Convoluting the natural line width, 23~MHz \cite{Jin93}, with the photon linewidth of $7.2\pm1.1$~MHz and the 4~MHz residual frequency fluctuations of the stabilized OPO results in an expected Lorentzian width of $34.2\pm1.1$~MHz, in very good agreement with the experimental finding. For comparison, Fig.~\ref{fig:spectrum} also shows an absorption line measured with a narrowband diode laser; its width of 24.6~MHz is very close to the natural line width. The maximum single-photon absorption rate on resonance, 690~s$^{-1}$ (taken from the blue curve in Fig.~\ref{fig:spectrum}), coincides with the measurement from the heralded absorption presented above. It is about three orders of magnitude larger than what has been reached before with Ca$^+$ \cite{Pir11, Huwer2013}.

\section{Summary and Conclusion}

In summary we presented an OPO-based, high-rate source of time-correlated, narrowband, widely tunable photon pairs interfacing the near infrared spectral region with the telecom bands. We demonstrated an essential building block of quantum networks, the heralded absorption of a single photon by a single atom; most importantly, the heralding photon is at a telecom wavelength, enabling long-range heralding in fiber networks. About $2.5 \cdot 10^6~{\rm s}^{-1}$ resonant pairs are generated in 7~MHz bandwidth; they effect an absorption rate close to $700~{\rm s}^{-1}$. In combination with single-photon frequency conversion steps \cite{Zas12} our SPDC source can also be used to establish quantum interfaces between dissimilar physical systems such as atoms or ions and solid-state qubits.

\begin{acknowledgments}
The work was funded by the German Federal Ministry of Science and Education (BMBF) within the projects "Q.com-Q" (contract No.\ 16KIS0127) and "QSCALE" (contract No.\ 01BQ1107). J.~Brito acknowledges support by CONICYT. 
\end{acknowledgments}


\begin{thebibliography}{ABC99}

\bibitem{Kimble2008}
H. J. Kimble, "The quantum internet", Nature {\bf 453}, 1023 (2008).

\bibitem{Duan2010} 
L.-M.~Duan, C.~Monroe, "Quantum networks with trapped ions", Rev. Mod. Phys. \textbf{82}, 1209-1224 (2010).

\bibitem{Zas12} 
S. Zaske, A. Lenhard, C. A. Ke\ss{}ler, J. Kettler, C. Hepp, C. Arend, R. Albrecht, W.-M. Schulz, M. Jetter, P. Michler, and C. Becher, "Visible-to-Telecom Quantum Frequency Conversion of Light from a Single Quantum Emitter," Phys. Rev. Lett. \textbf{109}, 147404 (2012).

\bibitem{Clausen2014}
C. Clausen, F. Bussi\`eres, A. Tiranov, H. Herrmann, C. Silberhorn, W. Sohler, M. Afzelius, and N. Gisin, "A source of polarization-entangled photon pairs interfacing quantum memories with telecom photons", New J. Phys. \textbf{16}, 093058 (2014).

\bibitem{Saglamyurek2015} 
E. Saglamyurek, J. Jin, V. B. Verma, M. D. Shaw, F. Marsili, S. W. Nam, D. Oblak, and W. Tittel, "Quantum storage of entangled telecom-wavelength photons in an erbium-doped optical fibre", Nature Photonics  \textbf{9}, 83 (2015).

\bibitem{Fek13} 
J.~Fekete, D.~Riel\"a{}nder, M.~Cristiani, and H.~de~Riedmatten, "Ultranarrow-Band Photon-Pair Source Compatible with Solid State Quantum Memories and Telecommunication Networks," Phys. Rev. Lett. \textbf{110}, 220502 (2013).

\bibitem{Fas04} 
S.~Fasel, O.~Alibart, S.~Tanzilli, P.~Baldi, A.~Beveratos, N.~Gisin, and H.~Zbinden, "High-quality asynchronous heralded single-photon source at telecom wavelength," New J. Phys. \textbf{6}, 163 (2004).

\bibitem{Pomarico2012}
E. Pomarico, B. Sanguinetti, T. Guerreiro, R. Thew, and H. Zbinden, "MHz rate and efficient synchronous heralding of single photons at telecom wavelengths", Optics Express, \textbf{20}, 23846.

\bibitem{Bao08} 
X.-H.~Bao, Y.~Qian, J.~Yang, H.~Zhang, Z.-B.~Chen, T.~Yang, and J.-W.~Pan, "Generation of Narrow-Band Polarization-Entangled Photon Pairs for Atomic Quantum Memories," Phys. Rev. Lett. \textbf{101}, 190501 (2008).

\bibitem{Haa09} 
A.~Haase, N.~Piro, J.~Eschner, and M.~W.~Mitchell, "Tunable narrowband entangled photon pair source for resonant single-photon single-atom interaction," Opt. Lett. \textbf{34}, 55-57 (2009).

\bibitem{Scholz2009}
M. Scholz, L. Koch, R. Ullmann, and O. Benson, "Single-mode operation of a high-brightness narrow-band single-photon source", Appl. Phys. Lett. \textbf{94}, 201105 (2009).

\bibitem{Wol11} 
F.~Wolfgramm, Y.~A.~de~Icaza~Astiz, F.~A.~Beduini, A.~Cerè, and M.~W.~Mitchell, "Atom-Resonant Heralded Single Photons by Interaction-Free Measurement," Phys. Rev. Lett. \textbf{106}, 053602 (2011).

\bibitem{Steinlechner2012}
F. Steinlechner, P. Trojek, M. Jofre, H. Weier, D. Perez, T. Jennewein, R. Ursin, J. Rarity, M. W. Mitchell, J. P. Torres, H. Weinfurter, and V. Pruneri, "A high-brightness source of polarization-entangled photons optimized for applications in free space", Optics Express \textbf{20}, 9640.

\bibitem{Pom09} 
E.~Pomarico, B.~Sanguinetti, N.~Gisin, R.~Thew, H.~Zbinden, G.~Schreiber, A.~Thomas, and W.~Sohler, "Waveguide-based OPO source of entangled photon pairs," New J. Phys. \textbf{11}, 113042 (2009).

\bibitem{Kra13} 
S.~Krapick, H.~Herrmann, V.~Quiring, B.~Brecht, H.~Suche, and Ch.~Silberhorn, "An efficient integrated two-color source for heralded single photons," New J. Phys. \textbf{15}, 033010 (2013).

\bibitem{Piro2009} 
N. Piro, A. Haase, M. Mitchell and J. Eschner, "An entangled photon source for resonant single-photon single-atom interaction", J. Phys. B: At. Mol. Opt. Phys. \textbf{42}, 114002 (2009).

\bibitem{Ou99} 
Z.~Y.~Ou, and Y.~J.~Lu, "Cavity Enhanced Spontaneous Parametric Down-Conversion for the Prolongation of Correlation Time between Conjugate Photons," Phys. Rev. Lett. \textbf{83}, 2556-2559 (1999).

\bibitem{Scholz2007}
M. Scholz, F. Wolfgramm, U. Herzog and O. Benson, "Narrow-band single photons from a single-resonant optical parametric oscillator far below threshold", Appl. Phys. Lett. \textbf{91}, 191104 (2007)

\bibitem{Monteiro2014}
F. Monteiro, A. Martin, B. Sanguinetti, H. Zbinden, and R. T. Thew, "Narrowband photon pair source for quantum networks", Opt. Express \textbf{22}, 4371 (2014).

%\bibitem{Ahl13} 
%A.~Ahlrichs, C.~Berkemeier, B.~Sprenger, and O.~Benson, "A monolithic polarization-independent frequency-filter system for filtering of photon pairs," Appl. Phys. Lett. \textbf{103}, 241110 (2013).

\bibitem{Schi13} 
P.~Schindler, D.~Nigg, T.~Monz, J.~T.~Barreiro, E.~Martinez, S.~X.~ Wang, S.~Quint, M.~F.~Brandl, V.~Nebendahl, C.~F.~Roos, M.~Chwalla, M.~Hennrich, and R.~Blatt, "A quantum information processor with trapped ions," New J. Phys. \textbf{15}, 123012 (2013).

\bibitem{Pir11} 
N.~Piro, F.~Rohde, C.~Schuck, M.~Almendros, J.~Huwer, J.~Ghosh, A.~Haase, M.~Hennrich, F.~Dubin, J.~Eschner, "Heralded single-photon absorption by a single atom", Nat. Phys. \textbf{7}, 17-20 (2011).

\bibitem{Huwer2013}
J. Huwer, J. Ghosh, N. Piro, M. Schug, F. Dubin and J. Eschner, "Photon entanglement detection by a single atom", New J. Phys. \textbf{15}, 025033 (2013). 
 
%Blatt_cavity
\bibitem{Stute2012}
A. Stute, B. Casabone, P. Schindler, T. Monz, P. O. Schmidt, B. Brandst\"atter, T. E. Northup, and R. Blatt, "Tunable ion-photon entanglement in an optical cavity", Nature \textbf{485}, 482 (2012). 

\bibitem{Kurz2014} 
C. Kurz, M. Schug, P. Eich, J. Huwer, P. M\"uller and J. Eschner, "Experimental protocol for high-fidelity heralded photon-to-atom quantum state transfer", Nat. Commun. \textbf{5}, 5527 (2014).

%latest_Blatt_cavity
\bibitem{Casabone2015}
B. Casabone, K. Friebe, B. Brandst\"atter, K. Sch\"uppert, R. Blatt, and T. E. Northup, "Enhanced quantum interface with collective ion-cavity coupling", Phys. Rev. Lett. \textbf{114}, 023602 (2015).

\bibitem{Moehring2007}
D. L. Moehring, P. Maunz, S. Olmschenk, K. C. Younge, D. N. Matsukevich, L.-M. Duan and C. Monroe, "Entanglement of single-atom quantum bits at a distance", Nature \textbf{449}, 68 (2007).

\bibitem{Schu13} 
M. Schug, J. Huwer, C. Kurz, P. M{\"u}ller, J. Eschner, "Heralded Photonic Interaction between Distant Single Ions", Phys. Rev. Lett. \textbf{110}, 213603 (2013).

%\bibitem{War14} C.~Warschburger, S.~Zaske, A.~Lenhard, M.~Schug, J.~Eschner, and C.~Becher, "Analysis of Frequency Noise Properties of a CW Optical Parametric Oscillator using an Optical Frequency Comb", Manuscript in preparation (2014).

%\bibitem{Mat13} Matthiesen, C. et al. "Phase-locked indistinguishable photons with synthesized waveforms from a solid-state source", Nat. Commun. 4:1600, doi: 10.1038/ncomms2601 (2013).

\bibitem{Zas11} 
S.~Zaske, A.~Lenhard, C.~Becher, "Efficient frequency downconversion at the single photon level from the red spectral range to the telecommunications C-band", Optics Exp. \textbf{19}, 12825-12836 (2011).

\bibitem{Zas10} 
S.~Zaske, D.-H.~Lee, and C.~Becher, "Green-pumped cw singly resonant optical parametric oscillator based on MgO:PPLN with frequency stabilization to an atomic resonance," Appl. Phys. B \textbf{98}, 729-735 (2010).

\bibitem{Mat12} 
C.~Matthiesen, A.~N.~Vamivakas, M.~Atat\"u{}re, "Subnatural Linewidth Single Photons from a Quantum Dot", Phys. Rev. Lett. \textbf{108}, 093602 (2012).

\bibitem{Ulh12} 
A.~Ulhaq, S.~Weiler, S.~M.~Ulrich, R.~Ro\ss{}bach, M.~Jetter, P.~Michler, "Cascaded single-photon emission from the Mollow triplet sidebands of a quantum dot", Nat. Phot. \textbf{6}, 238-242 (2012).

\bibitem{Cla11} 
C.~Clausen, I.~Usmani, F.~Bussieres, N.~Sangouard, M.~Afzelius, H.~de~Riedmatten, N.~Gisin, "Quantum storage of photonic entanglement in a crystal", Nature \textbf{469}, 508 (2011).

\bibitem{Her08} 
U.~Herzog, M.~Scholz, O.~Benson, "Theory of biphoton generation in a single-resonant optical parametric oscillator far below threshold", Phys. Rev. A \textbf{77}, 023826 (2008).

\bibitem{Piro2015}
N. Piro and J. Eschner, "Single photon absorption by a single atom: from heralded absorption to polarization state mapping", arxiv:1502.04349.

\bibitem{Sup} See Supplemental Material for details on the calculation of rates.

\bibitem{Jin93} J. Jin, and D. A. Church, "Precision Lifetimes for the Ca$^+$ 4p $^2$P Levels: Experiment Challenges Theory at the 1~\% Level," Phys. Rev. Lett. \textbf{70}, 3213-3216 (1993).

%\bibitem{Ure05} A.~B.~U’Ren, Ch.~Silberhorn, J.~L.~Ball, K.~Banaszek, I.~A.~Walmsley, "Characterization of the nonclassical nature of conditionally prepared single photons", Phys. Rev. A \textbf{72}, 021802(R) (2005).



\end{thebibliography}

\newpage
\clearpage

\section{Supplementary Material: Telecom-heralded single photon absorption by a single atom}
In this supplementary material we explain in more detail the method to derive the generated photon pair rate from the heralded absorption measurement including coincidence detection between quantum jumps and telecom photons.

\begin{figure}[htb]
	\includegraphics[width=0.45\textwidth]{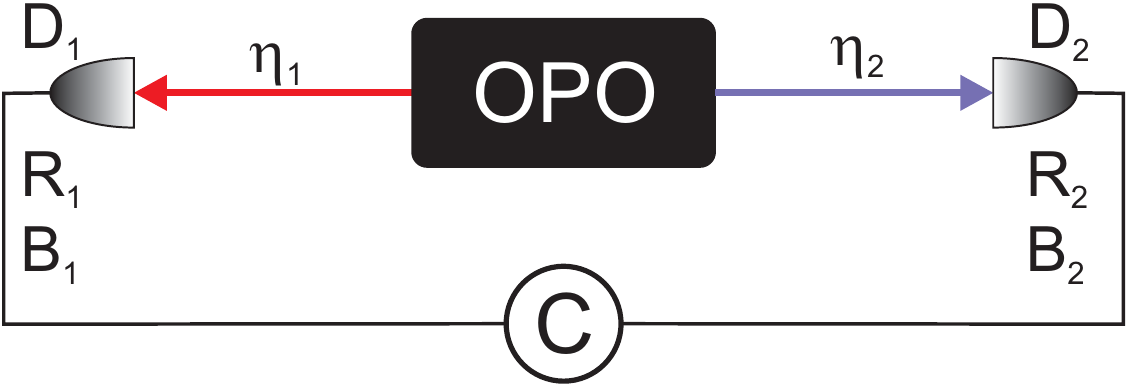}
	\caption{Schematic diagram of the setup indicating the definitions of rates and efficiencies. For details see text. \label{fig:SuppRates}}
\end{figure}

The OPO SPDC source produces photon pairs at a rate $P$, split in signal and idler mode. There is additional background noise generated which we account for by the rates $B_{1,2}$ (we hereby neglect background generated by the detectors in accordance with the experiment). Both signal and background photons suffer losses $\eta_{1,2}$ before being detected individually at detectors $D_{1,2}$. The factors $\eta_{1,2}$ include the detection efficiencies. The connection of these parameters is sketched in Fig.~\ref{fig:SuppRates}.
We can either detect singles-rates $R_{1,2}$ or measure the coincidence rate $C$ between signal and idler arm. This scheme also holds for quantum jump detection at the trapped ion in one of the arms.

In the paper we aim at deriving the generated pair rate $P$ and generated background rates $B_{1,2}$ from the measured single rates $R_{1,2}$ and coincidence rate $C$. Setting $\beta_{1,2}=B_{1,2}/P$, we identify the relations 
\begin{equation}R_{1,2} = (P+B_{1,2})\eta_{1,2} = P(1+\beta_{1,2})\eta_{1,2}\end{equation} 
and
\begin{equation}C=P\eta_1\eta_2\end{equation}

We use the measured $g^{(2)}$ function to derive $C$ (total counts in the coincidence peak divided by the total measurement time $T$), as well as the background rate per bin, $BG$ (average number of counts per bin of size $\Delta t$, divided by $T$). Consistency demands that $BG=R_1R_2\Delta t$. 

From the measured values $R_1$, $R_2$, and $C$, we cannot derive all unknowns, but if we take known values for $\eta_1$ and $\beta_1$, then 
\begin{equation}P = \frac{R_1}{\eta_1 (1+\beta_1)}\end{equation}
and 
\begin{equation}\eta_2 = \frac{C}{P \eta_1}\end{equation}
and finally 
\begin{equation}1+\beta_2 = \frac{R_2}{P \eta_2}\end{equation}

We apply this to the data of Fig.~5 using subscripts 1 for the quantum jumps and 2 for the SSPD counts. We measured
$$ R_1 = 111~{\rm s}^{-1} $$
$$ R_2 = 136,000~{\rm s}^{-1} $$
$$ C = \frac{16692}{310 \cdot 60~{\rm s}} = 0.9~{\rm s}^{-1} $$

and we suppose that we know the loss between OPO and ion (specified below), as well as the background at 854~nm, which we assume to be $B_1=0$. 

The calculation of $P$ is not done directly from Eq.~(3), because the ion is not immediately re-prepared after a quantum jump. This means further absorptions cannot be detected until the next preparation cycle. %We observe a quantum jump in nearly any of our preparation cycles (XX~\%) leading to a saturation effect detecting absorption events. Instead we get $P$ via the maximum observed absorption rate, derived from the exponential quantum jump probability per time. 
Instead we get $P$ via the maximum observed absorption rate. In analogy to the spectroscopy measurement we generate a histogram of the time distribution of quantum jumps and use this data to calculate the absorption rate via the effective lifetime (the procedure is explained in more detail in \cite{Schu13}). This rate is $R_{\rm abs} = 680~{\rm s}^{-1}$. The "dark time" after a quantum jump is equivalent to an additional efficiency factor, $\eta_{\rm sat}=R_1/R_{\rm abs}=0.163$, contributing to the loss on the ion side, which otherwise is given by 

\begin{equation} \eta_1 = \eta_{\rm SMF}\eta_{\rm fbs}\eta_{\rm ion}\eta_{\rm sat} = 4.4 \cdot 10^{-5} \end{equation} 

\noindent with measured values $\eta_{\rm SMF}=0.27$, $\eta_{\rm fbs}=0.5$, and $\eta_{\rm ion}=0.002$ for the fiber coupling and transmission loss between OPO and ion laboratory, the fiber beam splitter in front of the ion, and the absorption probability of the ion per arriving OPO photon, respectively. 

From this we get the estimated pair rate
\begin{equation} P = \frac{R_1}{\eta_1} = \frac{R_{\rm abs}}{\eta_{\rm SMF}\eta_{\rm fbs}\eta_{\rm ion}} = 2.52 \cdot 10^6~{\rm s}^{-1} \label{EstPairRate} \end{equation}  
and the other parameters according to Eqs.~(4,5) are
$$ \eta_2 = 8.1 \cdot 10^{-3} $$
and finally 
$$ \beta_2 = 5.7 $$

Regarding $\eta_2$ there are known contributions, the OPO to SSPD coupling and transmission loss $\eta_{\rm OPO-SSPD}=0.2$ and the SSPD detection efficiency $\eta_{\rm SSPD}=0.25$, such that there remains an unknown contribution
$$ \eta_{\rm unknown} = \frac{\eta_2}{\eta_{\rm OPO-SSPD}\eta_{\rm SSPD}} = 0.16 $$

Multiplying the estimated pair rate P, Eq.~(\ref{EstPairRate}), with the transmission loss $\eta_{\rm SMF}$ and $\eta_{\rm fbs}$ results in a rate of $\sim 340,000~{\rm s}^{-1}$ resonant photons available at the ion trap for experiments. This value is consistent with the results of the spectroscopy experiment, where we record about $8,000~{\rm s}^{-1}$ photons on the APD (with $\sim 0.3$ detection efficiency) after the filter cavities (with about 0.6 transmission and 0.13 coupling efficiency). Using an asymmetric rather than 50/50 beam splitter between ion and filter cavities would increase the rate of ion-resonant photons up to 2-fold.

\end{document}